\documentclass[12pt]{iopart}
\usepackage{graphicx}
\begin{document}

\title{ 
Monte Carlo Study of Mixed-Spin S=(1/2,1) Ising Ferrimagnets
}
\author{ W Selke$^{1,2}$ and J Oitmaa$^2$}
\address{$^1$ Institut f\"ur Theoretische Physik, RWTH Aachen, 52056
  Aachen, Germany}
\address{$^2$ School of Physics, The University of New South Wales,
  Sydney, NSW 2052, Australia}
\date{\today}
\begin{abstract}
We investigate Ising ferrimagnets on square and
simple--cubic lattices with exchange couplings between
spins of values S=1/2 and S=1 on 
neighbouring sites  and an additional  single--site anisotropy term
on the S=1 sites. Based mainly on a careful and comprehensive Monte Carlo
study, we conclude that there is
no tricritical point in the two--dimensional case, in contradiction
to mean-field predictions and recent series results. However, evidence for a
tricritical point is found in the three--dimensional case. In addition, a
line of compensation points is
found for the simple--cubic, but not for the square lattice.
\end{abstract}
\pacs{75.10.-b, 75.10.Hk, 75.40.Mg, 75.50.Gg}
\submitted{\JPCM}

\section{Introduction}
\label{sec1}

Mixed-spin Ising models have been studied for some time as simple models
of ferrimagnets, and there has been renewed interest recently in
connection with 'compensation points'. These are temperatures, below the
critical temperature, at which the sublattice magnetizations cancel
exactly, giving zero total moment. As the temperature is tuned through
such a point the total magnetization changes sign, which may be
used in technological applications. In this context,
Ising models may be exactly solvable in special cases \cite{gon,lip,jas1,dak}
 or they may be  studied by
a variety of powerful approaches, including 
Monte Carlo \cite{oit1,jas2,zha,bue,nak,gol} or
other \cite{god,oit2,kan,boe,sar} methods.
In the present work we revisit one of the simplest such models, a mixed--spin
Ising model with spins S=1/2 and 1 occupying the sites of a bipartite
square or simple cubic lattice with the Hamiltonian

\begin{equation}
{\cal H} = -J \sum_{\langle i,j\rangle} {\sigma_i S_j} + D \sum_{j\in B} {S_j}^2
\end{equation}

with couplings $J$ between 
spins $\sigma_i = \pm 1$ on the sites of sublattice 'A', and neighbouring 
spins $S_j = 1,0,-1$ on sites forming the sublattice 'B'. D denotes the
strength of a single--ion term acting only on the S=1 spins of sublattice B.
Following previous convention, we
choose $\sigma_i = \pm 1$ rather than $\pm 1/2$, which has
to be taken into account when calculating sublattice
magnetizations and when defining the compensation
point. The convention simply amounts to a rescaling of
the exchange coupling. Note that the nearest neighbour coupling $J$
may be either antiferromagnetic, $J<0$, as assumed often for 
ferrimagnets, or ferromagnetic, $J>0$. Both cases are completely equivalent
by a simple spin reversal on either sublattice. We shall use in
this article ferromagnetic couplings. As a
consequence, in our case the magnetizations of both sublattices
are identical at the compensation point, while in the
antiferromagnetic case, at the same compensation point, the
sublattice magnetizations have equal magnitude but different sign
leading to the above mentioned vanishing of the total magnetization.

The model on the square lattice has been studied by several
authors. Kaneyoshi and Chen \cite{kan}, via a
mean-field treatment, found a line of
compensation points in
a narrow region $ 4 > D/J \ge 2$ ln6 (= 3.583..) and a tricritical
point at $D_t/J = 3.72$, i.e. a first-order transition for $D > D_t$. 
Buendia and Novotny \cite{bue}, using transfer matrix methods, supplemented 
by Monte Carlo simulations, found no evidence of either a
compensation point or a tricritical point, although a
compensation point was observed in an extended model
with additional ferromagnetic interactions between $\sigma$
spins. More recently, Oitmaa and Enting \cite{oit3} studied the same model
using a combination of high- and low-temperature series. No compensation
point was found, but evidence for a first-order transition, and hence a
tricritical point was observed from an apparent crossing of the high- and
low-temperature branches of the free energy with different slopes, for
$D/J \ge 3.2$. Thus the phase diagram of this simple model remained
uncertain, motivating partly the present extensive
Monte Carlo study, improving previous simulations substantially. In
fact, our study provides
clear evidence that the model in two dimensions has no
compensation point or tricritical point. Moreover, the
model is found to exhibit very interesting thermal behaviour, both for the
specific heat and the magnetization, especially in the
low--temperature region near $D = 4$, which has not been discussed in
detail before. This behaviour is the likely explanation for the
apparent 'first-order' behaviour observed in Ref. 16.

For the simple cubic lattice, to our knowledge, no detailed analyses 
have been done so far. Of course, mean--field theory may be
easily applied, leading again to a tricritical point and a line of
compensation points. 

The outline of the article is as follows. In Section 2 we present and
discuss our results for the square lattice. In Section 3 we
consider the simple--cubic lattice. Here, in contrast to the
two--dimensional case, we find a clear occurrence of a line
of compensation points. Furthermore, we
obtain clear evidence of transitions of first-order, and thence
of a tricritical point, which we
locate approximately. In the final section, a brief summary
will be given.

\section{The model on the square lattice}
\label{sec2}
 
Let us first consider the ferrimagnet, eq. (1), in the case of
a square lattice. We have performed mainly standard Monte Carlo
simulations, using the Metropolis algorithm with
single--spin flips, providing, indeed, the required
accuracy,  so that there was no need to apply other techniques
like cluster--updates or the Wang--Landau approach \cite{bil}. We studied
lattices with $L$x$L$ sites, employing full
periodic boundary conditions. $L$ ranged 
from 4 to 80, to study finite--size effects. Typically, runs
of $10^7$ Monte Carlo steps per spin have been done, with
averages and error bars obtained from evaluating a number of such
runs, at least three, using different random numbers.
These rather long runs lead to very good statistics, improving appreciably
results of previous simulations \cite{zha,bue}. The estimated errors, unless
shown otherwise, are smaller than the symbols depicted in the
figures.

\begin{figure} [h]
\vspace{1cm}
\begin{center}
 \includegraphics[width=0.55\linewidth]{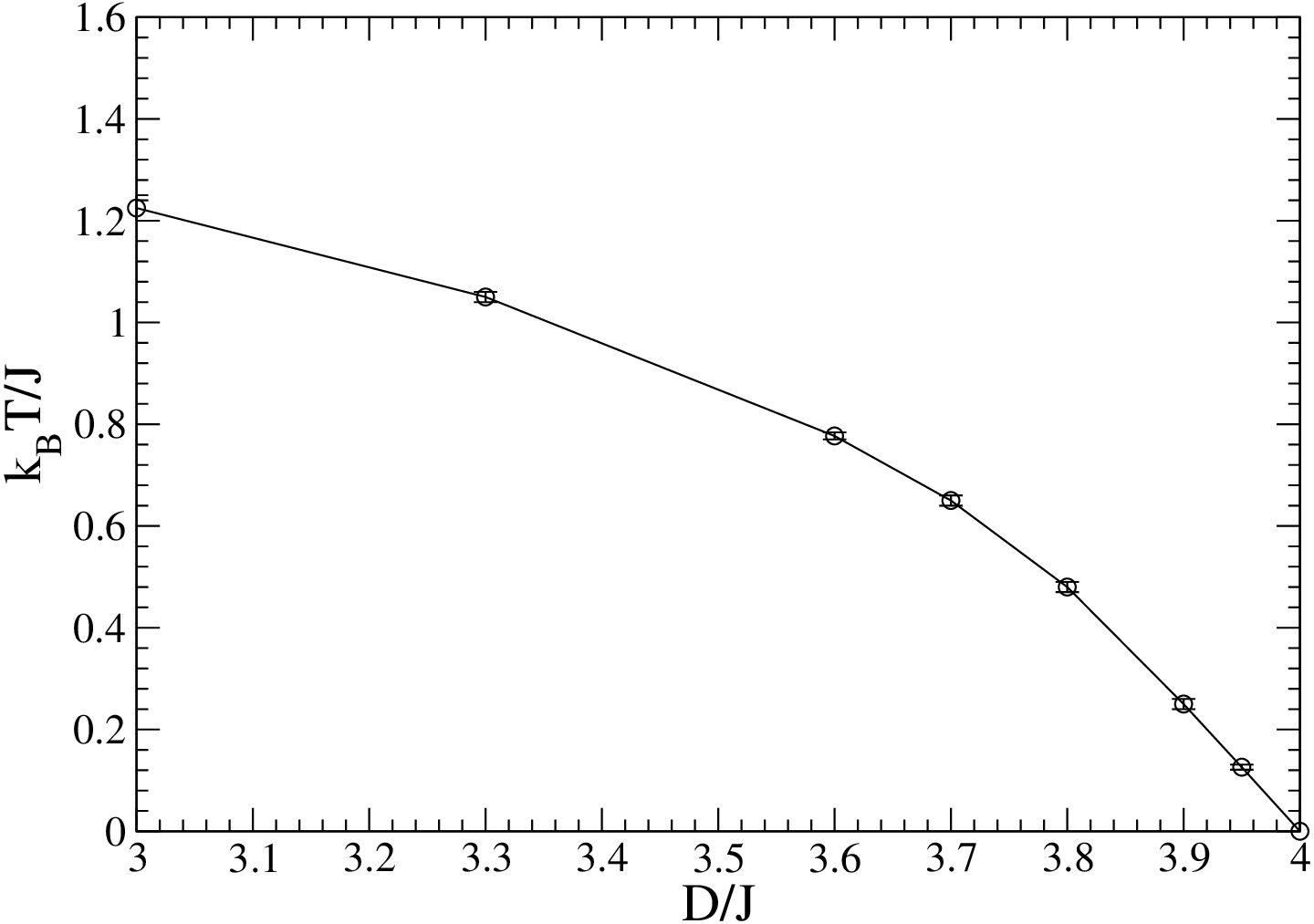}
\end{center}
\label{fig1}
\vspace{-0.5cm}
\caption{Phase diagram of the mixed--spin model on a square lattice.} 
\end{figure}

We recorded the energy per site, $E$, the
specific heat, $C$, both from the energy fluctuations 
and from differentiating $E$ with respect to the temperature, and
the absolute values of the sublattice magnetizations
of the two sublattices

\begin{equation}
|m_A| = <|\sum_{A} {\sigma_i}|>/(2 (L^2/2))
\end{equation}

and 

\begin{equation}
|m_B| = <|\sum_{B} {S_j}|>/(L^2/2)
\end{equation}

as well as the absolute value of the total magnetization,

\begin{equation}
|m| = <|\sum_{A} {\sigma_i} + \sum_{B}{S_j}|>/L^2
\end{equation}

where the brackets $< >$ denote the thermal average. Note the factor
of 1/2 in the definition of $|m_A|$, taking into account the
correct length of the S=1/2 spins, so that $|m_A(T=0)|= 1/2$,
while $|m_B(T=0)|= 1$ for the ferromagnetic ground state. In
addition, the corresponding
susceptibilities, $\chi_A$, $\chi_B$, and $\chi$, have been computed
from the fluctuations of the magnetizations. We also analysed 
histograms for the total magnetization, $p(m)$, i.e. the probability
to encounter a configuration with the magnetization $m$, as 
well as the fourth--order cumulant of
the order parameter, the Binder cumulant \cite{bin}, defined by

 \begin{equation}
  U = 1- <m^4>/(3 <m^2>^2)
\end{equation}

with $<m^2>$ and $<m^4>$ being the second and fourth moment of the 
total magnetization. Finally, we monitored typical equilibrium
Monte Carlo configurations, illustrating the microscopic
behaviour of the system.

To test the accuracy of the simulations, we computed numerically
exact results for various quantities by enumerating all possible
configurations for small lattices with $L= 4$. 

\begin{figure}[h]
\vspace{1cm}
\hspace{-1.2cm}
\begin{center}
\includegraphics[width=0.44\linewidth]{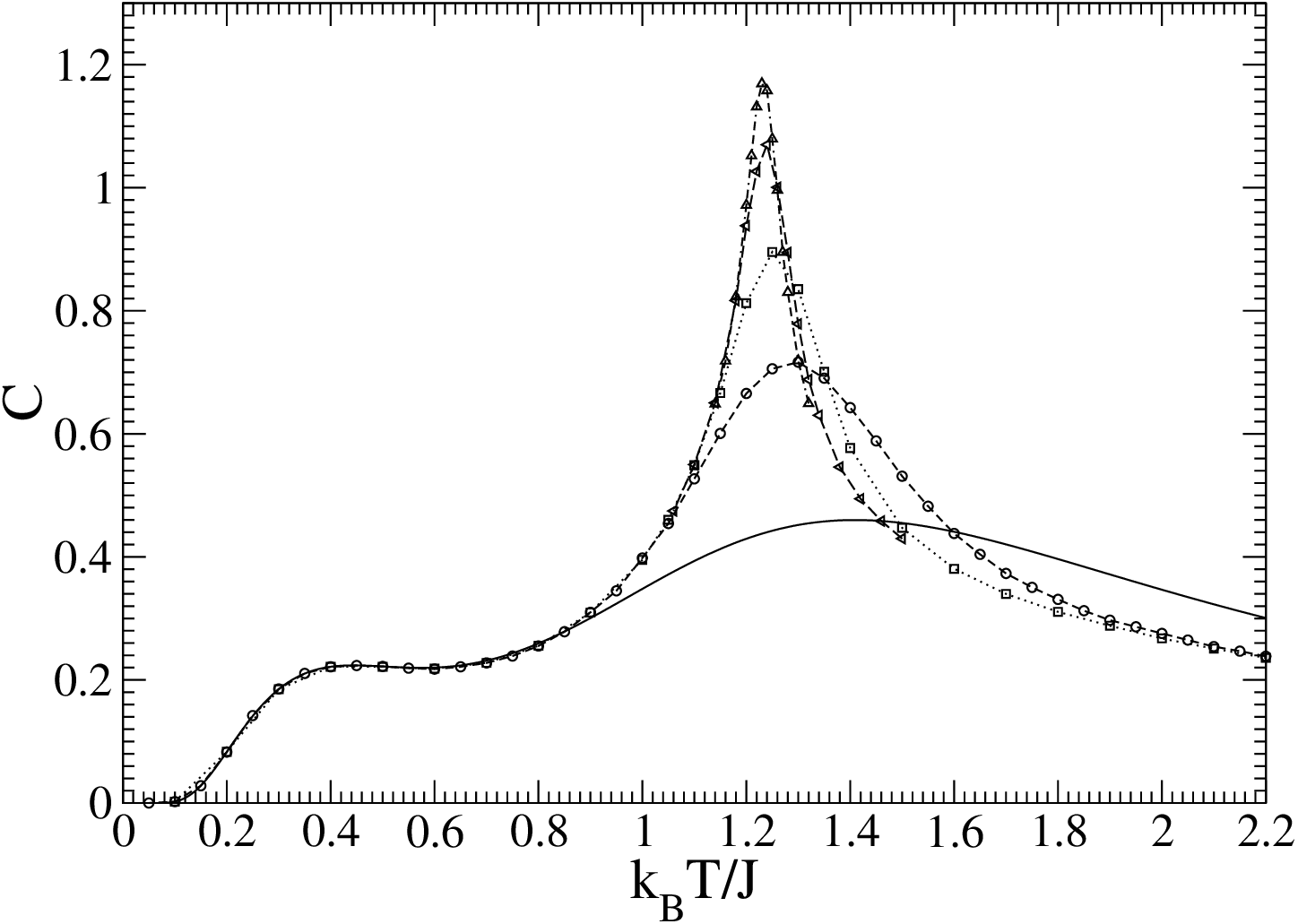}
\hspace{0.7cm}
\includegraphics[width=0.44\linewidth]{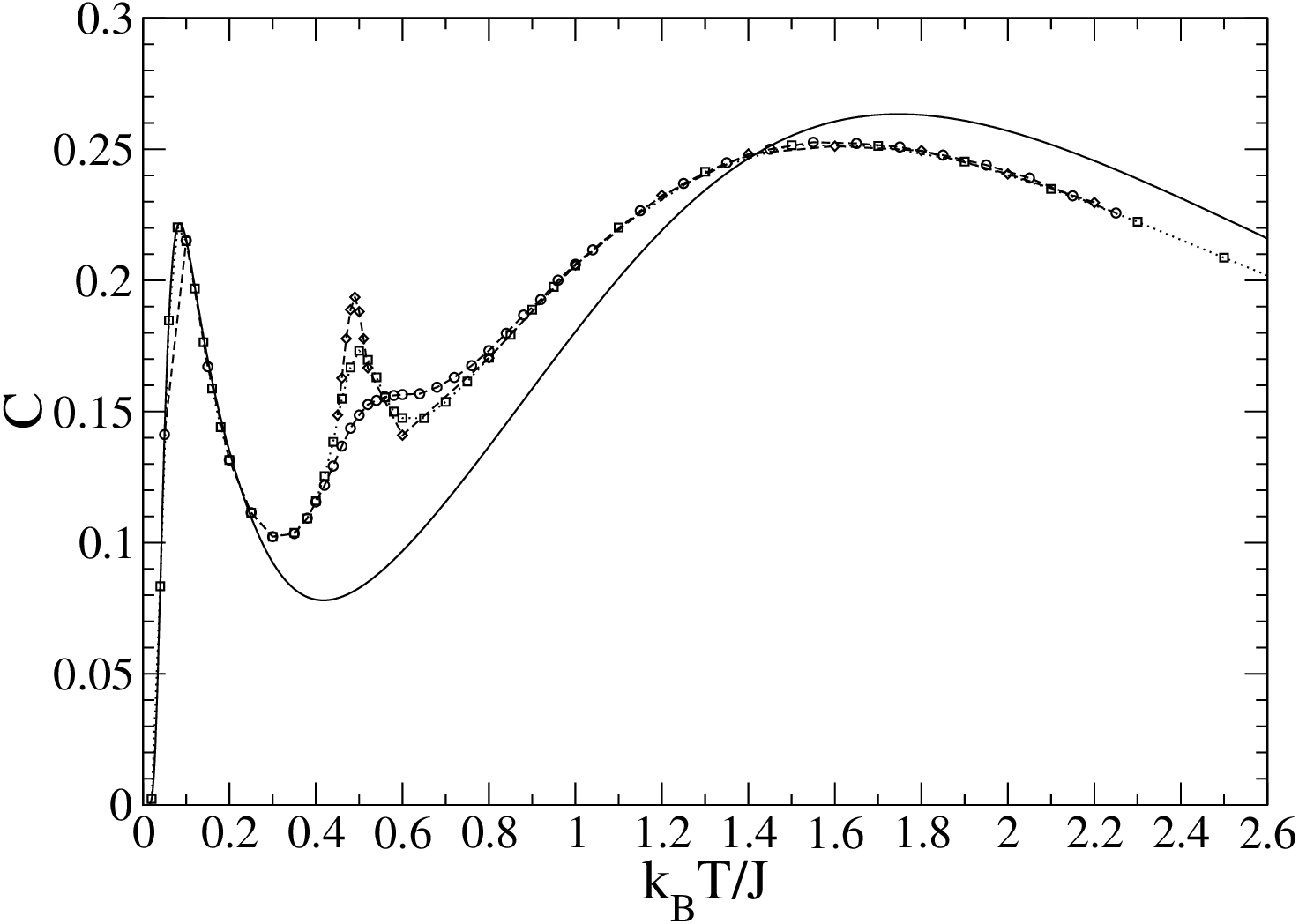}
\end{center}
\label{fig2}
\caption{ (a) Left: Specific heat at $D/J$= 3.0, showing
numerically exact, for $L=4$(solid line), and
Monte Carlo data for sizes $L$= 10 (circles), 20 (squares), 40
(diamonds) and 60 (triangles). (b) Right: Specific heat at
$D/J$=3.8, showing numerically exact, for 
 $L= 4$ (solid line), and Monte Carlo data for sizes $L$= 20 (circles),
 40 (squares) and 80 (diamonds).}
\end{figure}

In agreement with previous work, the model is observed
to display a ferromagnetic ground state and low--temperature
phase for $D/J <4$. The energy to flip a B spin from its ferromagnetic
orientation, '+' or '$-$', surrounded by four
A spins of the same orientation, to the state 0 is obviously
$\Delta E = 4J - D$ which vanishes at $D= 4J$. Hence the
ground state at $D/J= 4$ will comprise configurations with
'0' states on B sites and arbitrarily oriented spins
on the neighbouring A sites, as well as ferromagnetic plaquettes (of
either sign) on B sites and neighbouring A sites. Due to the
resulting high degeneracy, one
may call ($D/J=4, T=0$) the 'degeneracy point'. For $D > 4J$ at zero
temperature, all B spins will be in the state 0, with
the A spins being randomly oriented. This leads to a
lower, but still macroscopic degeneracy. At $D/J \ge 4$, there
is no ordered phase even at zero temperature.

Most of our Monte Carlo work deals with the interesting range
$3 \le D/J < 4$, which had been discussed controversially before,
augmented by some simulations at lower values of $D/J$. The
resulting phase diagram is depicted
in Fig. 1, based on monitoring the size--dependence
of the position of
the (critical) maxima in the specific heat and susceptibility, and
the intersection points
of the Binder cumulant, see below. Our findings are 
in accordance with a continuous
transition in the Ising universality class for all values of $D/J$ we
studied, $D/J \le 3.98$. There is no compensation point.

In the following, we shall discuss main properties of the
physical quantities mentioned above.

The specific heat $C$, for negative or relatively small positive
$D/J$, is observed to resemble qualitatively that of the nearest--neighbour
Ising model on a square lattice. There is a unique maximum
in $C(T)$, for finite
$L$, turning into a logarithmic singularity in the
thermodynamic limit. Indeed, in the
limit $D/J \rightarrow -\infty$, one recovers the simple
Ising model. Increasing $D/J$, as displayed in
Fig. 2a for $D/J= 3.0$, an additional shoulder or maximum evolves
at a lower temperature, $T_l$, being largely independent of
lattice size and being non--critical. Its origin becomes
clear by further increasing $D/J$, as shown in Fig. 2b for $D/J= 3.8$. In
fact, one finds $k_BT_l/J \approx 0.42 (4-D/J)$, reflecting
the thermally activated flipping of B spins from the 
ferromagnetic state '1' (or --'1') to the state zero, requiring, as
stated above, an energy proportional to $4-D/J$. It is interesting
to note that the height of the pronounced non--critical
peak, signalling
the partial disordering of the B sublattice, depends only very weakly on 
$D/J$. In the range $3.5 \le D/J < 4$, one has $C(T_l) \approx 0.22$.

As illustrated in Fig. 3b for $D/J= 3.8$, the critical peak, located
at $T_m$, may separate from the upper maximum, at
$T_u$, when increasing the strength of the single--ion 
term. Thus, the specific heat may display a three--peak
structure, with two non--critical maxima and a critical
peak in between. The origin of
the maximum at $T_u$ is due to the fact that at the critical
point, the $\sigma$ spins on the A sublattice form rather
large clusters of different orientations, leading to
the vanishing of the order parameter. That behaviour may be
seen by monitoring typical equilibrium configurations. These
clusters shrink quickly near $T_u$, due to thermally activated
flipping of $\sigma$ spins, determined by the coupling
constant $J$. Indeed, $T_u$ is essentially independent of $D$. As
seen in Fig. 3b, the maximum in $C$ at $T_u$ depends rather weakly on
the size of the lattice, $L$, demonstrating its non--critical character.    

The height of the critical maximum at $T_m$ is expected, for
Ising universality, to
increase logarithmically with $L$ for sufficiently large values
of $L$. Our results are consistent with this expectation. However, on
approach to the degeneracy point, the
background contribution to the specific heat becomes
more and more relevant. Then larger and larger lattices, with
$L > L_0$, are needed to see the anticipated logarithmic
behaviour. For example, at $D/J= 3.6$, one gets $L_0 \approx 40$, and
$L_0 \approx 60$ at $D/J=3.95$. In fact, in that range, the Ising--like
character of the transition may be inferred more clearly from
other quantities, as discussed below.

\begin{figure} [h]
\vspace{1cm}
\begin{center}
\includegraphics[width=0.55\linewidth] {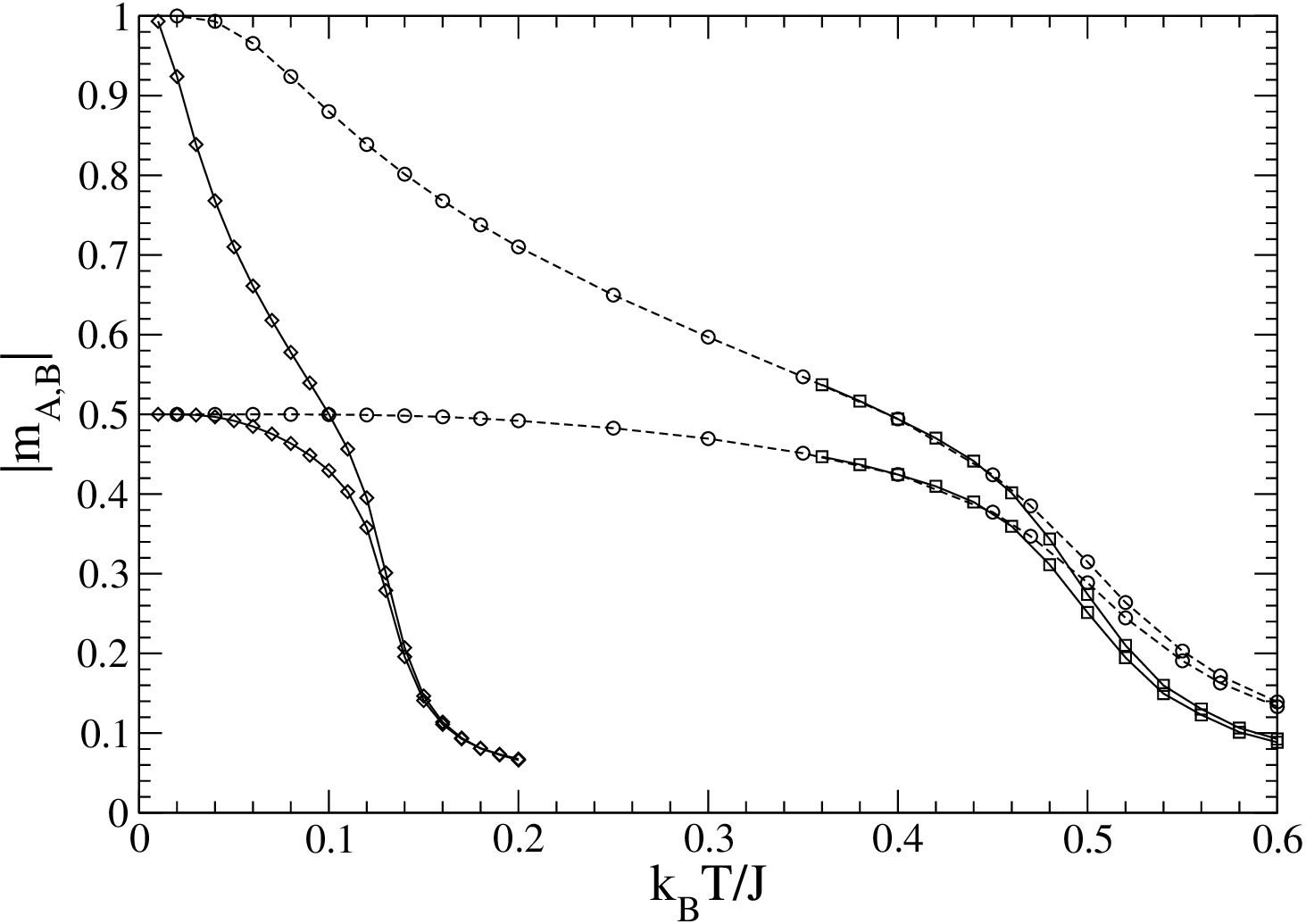}
\end{center}
\label{fig3}
\caption{Sublattice magnetizations $|m_A|$ and $|m_B|$ at $D/J= 3.8$
    (circles) and 3.95 (diamonds), with lattices
    of size $L$= 40 (dashed lines) and 60 (solid lines).} 
\end{figure}

The partial disordering of the B sublattice, near $T_l$, leads
to a rapid decrease of the magnetization $|m_B|$, as
illustrated in Fig. 3. Actually, the anomaly in $|m_B|$ becomes
more and more dramatic on approach to the degeneracy
point. In contrast, the magnetization of
the A sublattice, $|m_A|$, is hardly affected by
the disordering of the B sublattice. Indeed, this behaviour
may open the possibilty of a
compensation point, at which the two sublattice
magnetizations, $|m_A|$ and $|m_B|$, would coincide. However, as
depicted in Fig. 3, we find no evidence for such
a compensation point in two dimensions for all cases
we studied, with $D/J$ going up to 3.95. 

The susceptibility $\chi$ is found to show, in
all cases we studied, only one maximum, close to the critical
temperature. The background term is much weaker than
for the specific heat, allowing an analysis of critical 
properties for smaller lattices. In fact, as illustrated
in Fig. 4, the size dependence of the height of the maximum
in $\chi$, $\chi_{max}(L)$, is observed to be nicely compatible with
the asymptotic form $\chi_{max} \propto L^{7/4}$, expected
for the Ising universality class, for all
cases studied and sufficiently large lattices. Note that the
susceptibility shows a rather mild anomaly near
$T_l$, where the specific heat shows a pronounced maximum, close
to the degeneracy point. At
that anomaly, $\chi(T)$ exhibits
a maximal slope, as may be easily identified using 
exact enumeration for small lattices. The shrinking of
the A clusters, as indicated by the broad maximum in $C$
at $T_u$, leads to no obviously unusual features in
the susceptibility.

As usual, one may estimate the bulk transition temperature, $T_c$, from
the size dependent position of the corresponding peaks in $\chi$ and
$C$. We obtain consistent estimates, shown in Fig. 1, with the location
of the maxima varying, for large $L$, proportionally to $1/L$, as
expected for Ising--like transitions. Of course, one
gets distinct proportionality factors for the two quantities.

\begin{figure} [h]
\vspace{1cm}
\begin{center}
\includegraphics[width=0.55\linewidth]{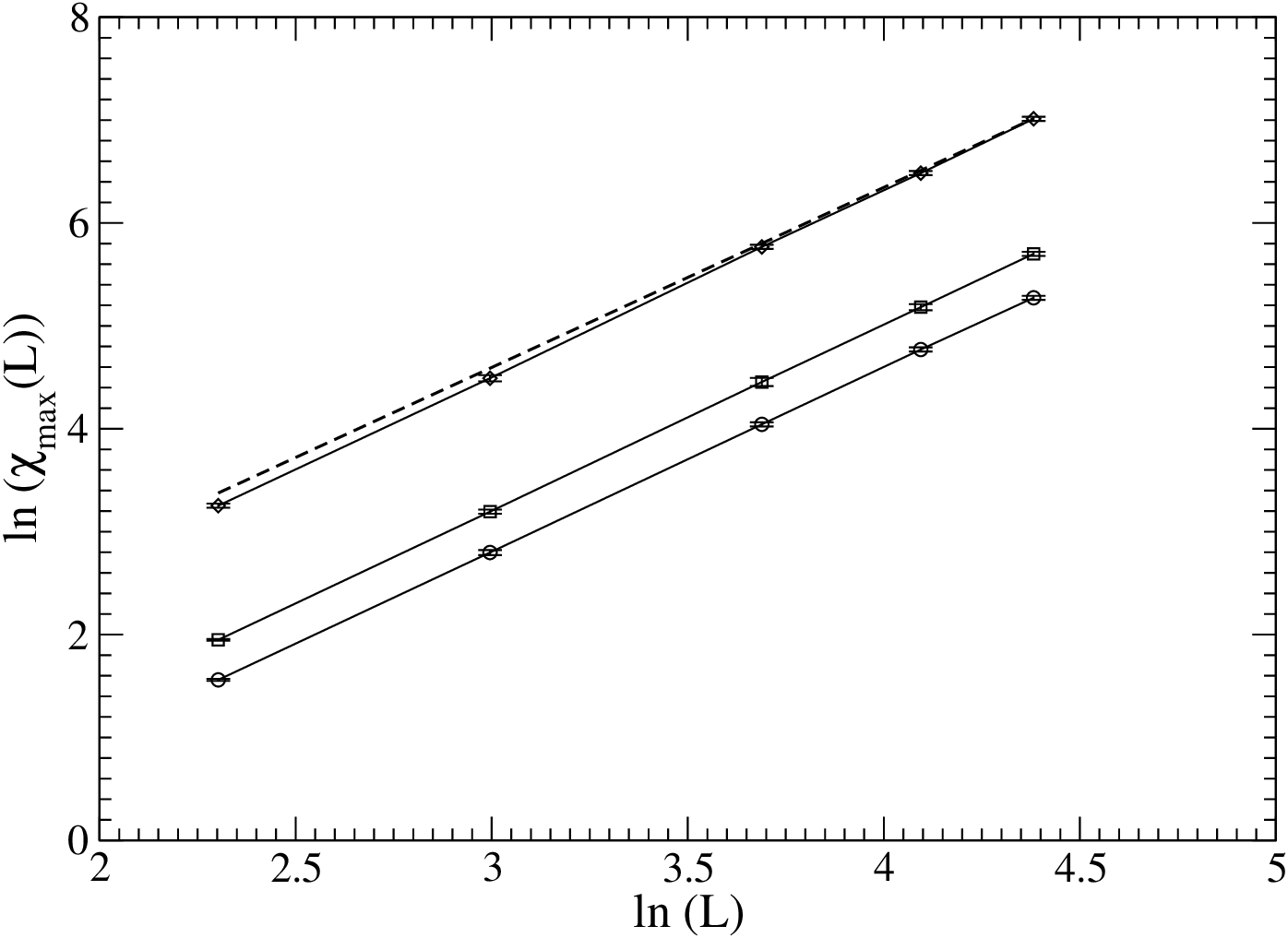}
\end{center}
\label{fig4}
\caption{Log--log plot of the susceptibility
 $\chi_{max}$ versus system size $L$ for the square lattice
 at D/J= 3.6 (circles), 3.8 (squares) and 3.95 (diamonds). For
 comparison, the dashed line shows $\chi_{max} \propto L^{7/4}$.} 
\end{figure}

The transition temperature may be also conveniently estimated from
the Binder cumulant, $U$. Indeed, the estimates follow
from the location of the intersection temperatures of the
cumulants for different lattice sizes \cite{bin}. Finite size
corrections often turn out to be rather small. Actually, this is
also true for the present model, as shown in Fig. 5 for
$D/J= 3.95$. We
find very good agreement with the estimates of
$T_c$ based on the susceptibilty and the
specific heat. Note
that the value of $U$ at the intersection temperature is, already
for fairly small systems sizes, close to the accurately known \cite{blo}
critical Binder
cumulant $U^*= U(T_c, L=\infty)$ for isotropic Ising
models, $U^*= 0.6069...$. One
may emphasize that anisotropic interactions and
correlations may lead to non--trivial dependences of $U^*$
on such interactions \cite{doh,sel}. However, here we are
dealing with an isotropic system, and excellent agreement with the
known critical value is observed, demonstrating that the
the transition belongs to the Ising universality class.

\begin{figure} [h]
\vspace{1cm}
\begin{center}
\includegraphics[width=0.55\linewidth] {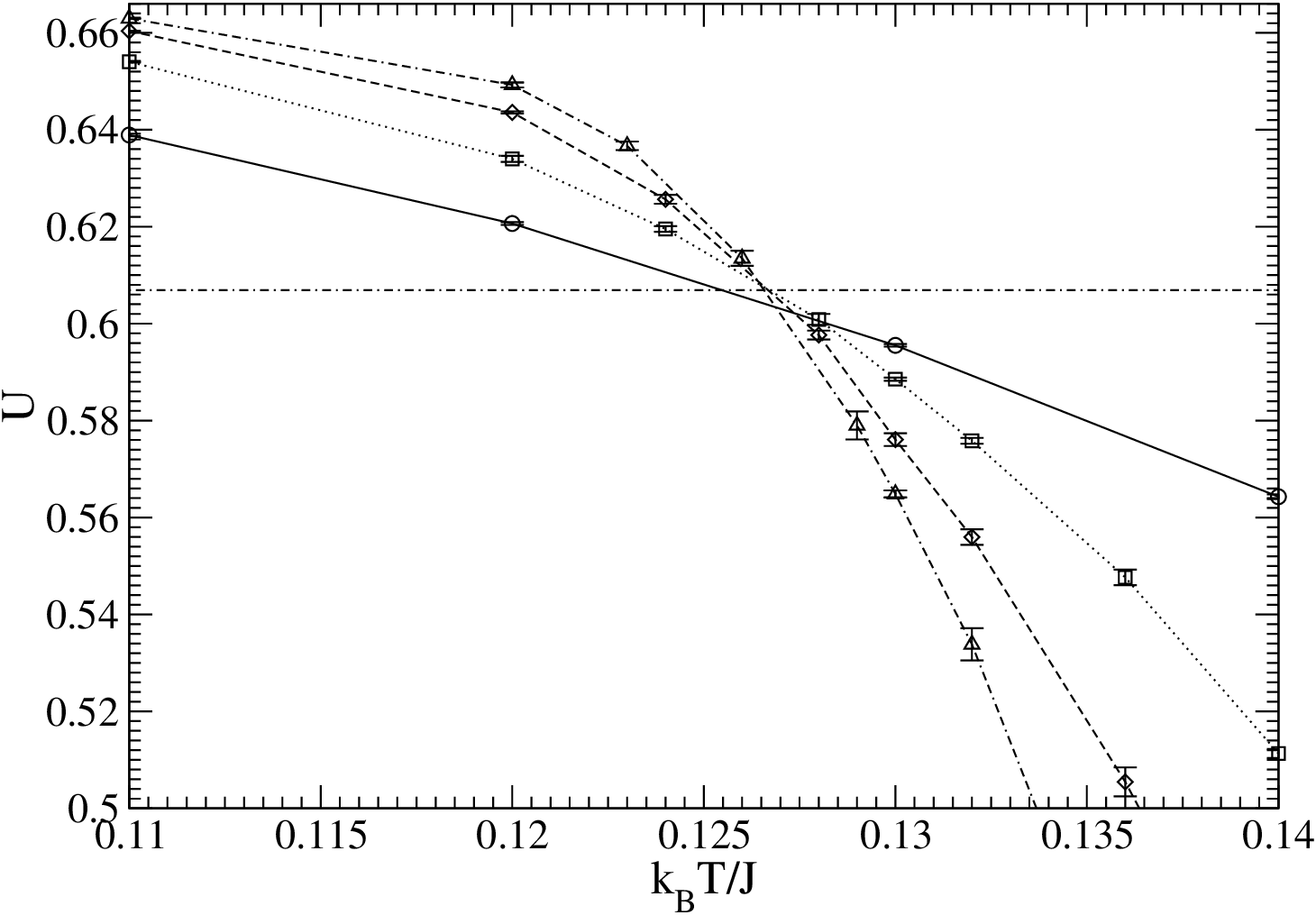}
\end{center}
\label{fig5}
\caption{ Binder cumulant $U(L,T)$ at D/J=3.95 for $L= $20 (circles),
    40 (squares), 60 (diamonds), and 80 (triangles). The horizontal
    line indicates the critical Binder cumulant of an isotropic Ising
    model in the thermodynamic
    limit \cite{blo}.} 
\end{figure}

Additional insight into the phase transition is provided
by the histograms for the total 
magnetization, $p(m)$. An example is displayed in Fig. 6. As
expected for a continuous transition, $p(m)$ shows, in
the ferromagnetic low--temperature phase, two symmetric
peaks, at $\pm m_0$, moving closer and closer to each other
on approach to $T_c$ and when increasing the lattice size. Above
$T_c$, $p(m)$ tends to acquire a Gaussian shape \cite {bin}. We
emphasize that Fig.6 refers to the case $D/J= 3.98$, i.e. very
close to the degeneracy point. There is no indication of
a transition of first order, which might be signalled by a
central peak, in addition to the two peaks at $\pm m_0$, as would be
the case for coexistence of the disordered and ordered
phases. Accordingly, we may safely conclude, based
on the analysis of several quantities, that we have clear
evidence for continuous transitions of Ising type
along the boundary of the ferromagnetic phase, at least
for the region $D/J \le 3.98$.

\begin{figure} [h]
\vspace{1cm}
\begin{center}
\includegraphics[width=0.55\linewidth] {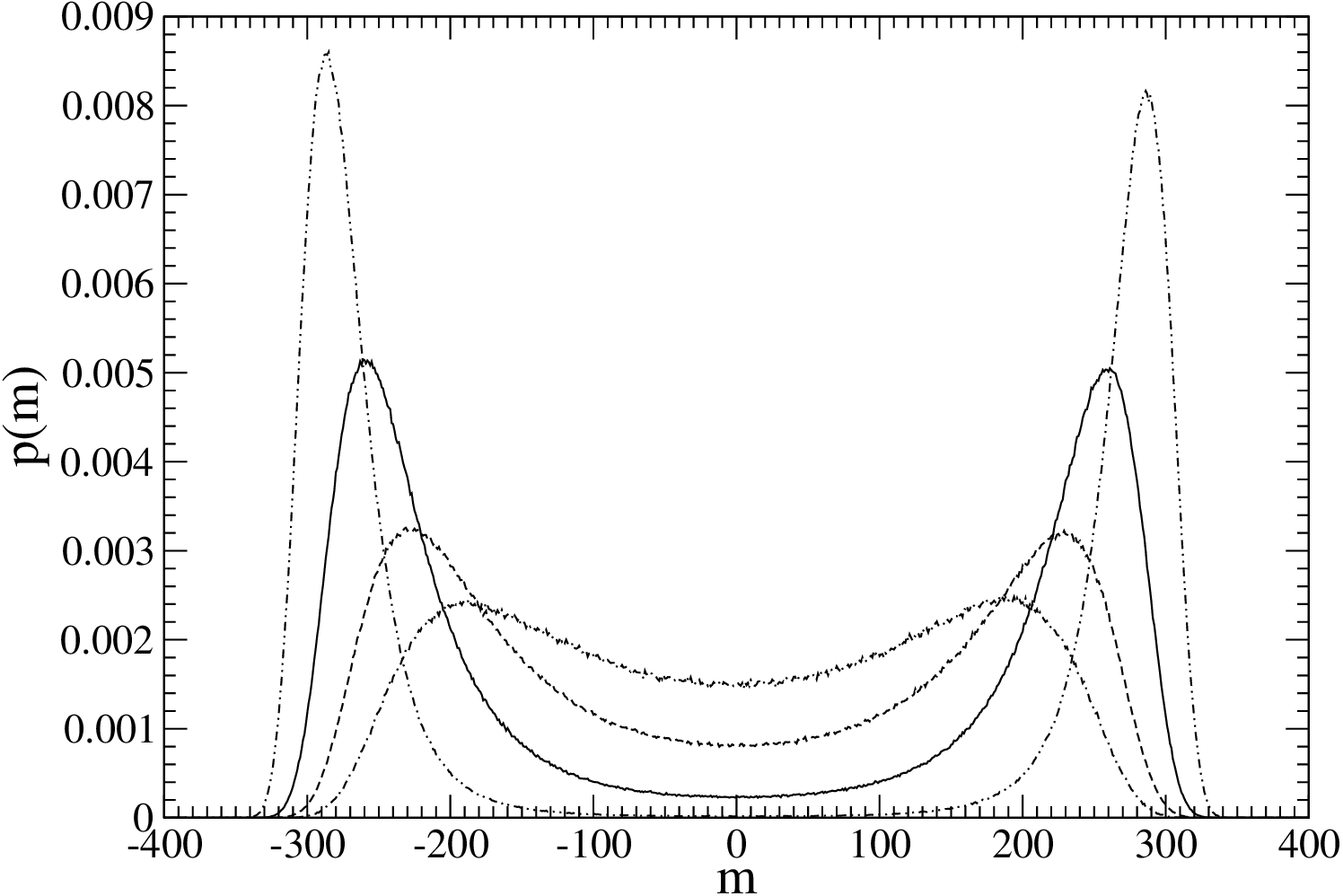}
\end{center}
\label{fig6}
\caption{Histogram of the total magnetization, $p(m)$, for the
 square lattice with $L= 20$ at $D/J= 3.98$ and temperatures below and above 
    the transition, $k_BT/J= 0.04, 0.05, 0.06$, and 0.07 (peak
    positions moving towards the center), with $k_BT_c/J \approx 0.051$.} 
\end{figure}

\section{The model on the simple--cubic lattice}
\label{sec3}

Let us now turn to the analysis of the mixed--spin model, eq. (1), on
a simple cubic lattice. In complete analogy to the two--dimensional
case, we did standard Monte Carlo simulations, applying
the Metropolis algorithm. We studied lattices with $L^3$ sites, with
$L$ ranging
from 4 to 32. Full periodic boundary conditions were
employed. Typically, runs of $2\times10^6$ to $5\times10^6$ Monte Carlo steps
per spin were performed, averaging over a few, at least three, such runs to
estimate thermal averages and error bars.

\begin{figure} [h]
\vspace{1cm}
\begin{center}
\includegraphics[width=0.55\linewidth] {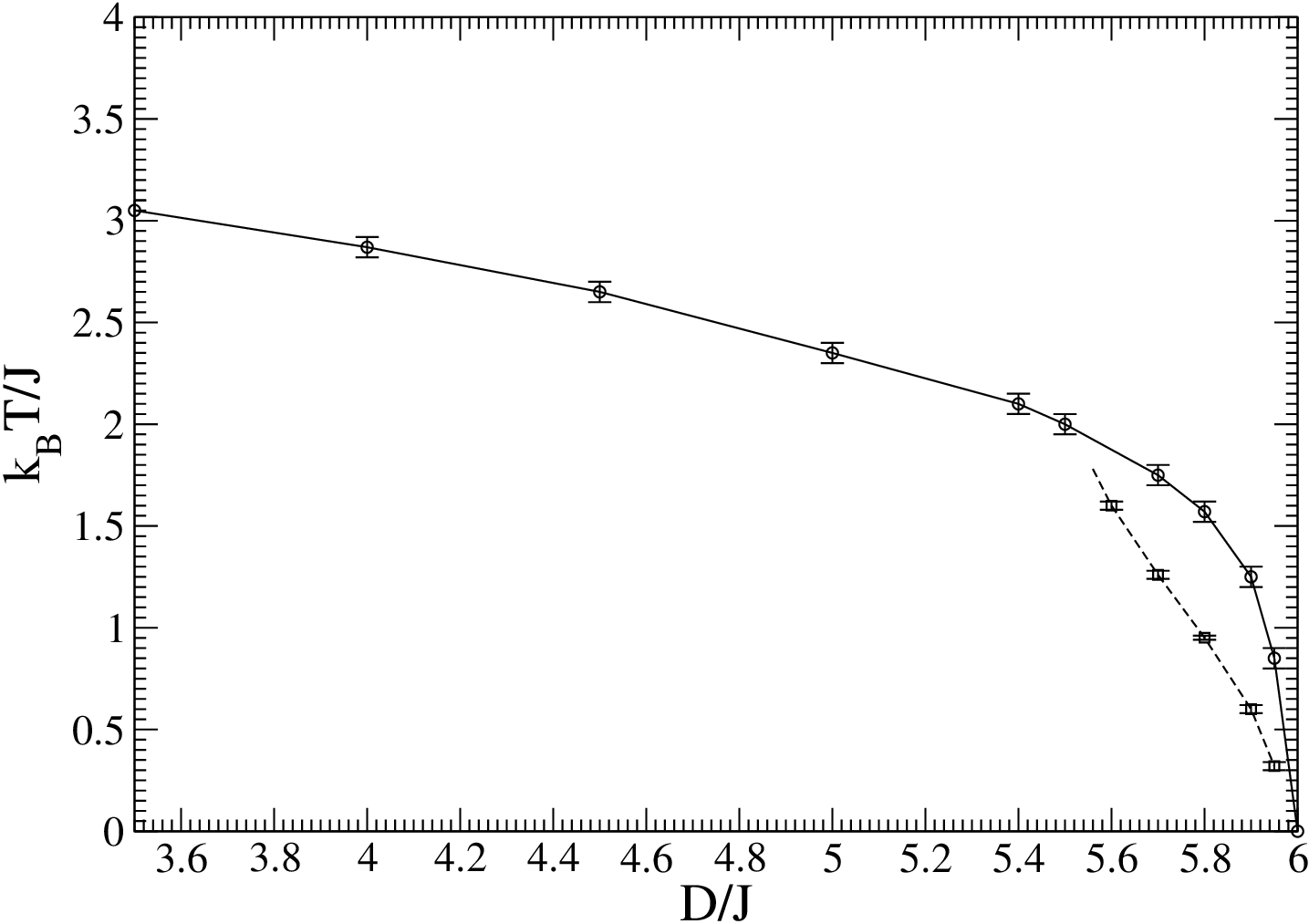}
\end{center}
\label{fig7}
\caption {Phase diagram of the mixed--spin model on a simple--cubic
lattice. The solid line denotes the boundary of the ferromagnetic
phase, while the dashed line denotes compensation points.} 
\end{figure}

As for the square lattice, the energy $E$, the specific heat
$C$, magnetizations $|m_A|, |m_B|$, and $|m|$, as well as corresponding
susceptibilities, the Binder cumulant $U$, and histograms
for the total magnetization, $p(m)$, were
recorded. Typical Monte Carlo equilibrium configurations were generated 
to illustrate the microscopic behaviour.

For the cubic lattice, one has a ferromagnetic ground 
state at $D/J < 6$. The degeneracy point occurs now at $D/J= 6$, with
ground states comprising local ferromagnetic
plaquettes of neighbouring A and B spins as well
as B spins in the state 0 with surrounding
A spins being randomly oriented. For $D/J >6$, a high, but
reduced degeneracy prevails, with all B spins being zero, and the
A spins pointing randomly 'up' or 'down'. 

For $D/J$ small or negative, a continuous transition
of Ising type is expected to occur, as we confirm in
simulations with moderate efforts. Most of our work has
been done for $3.5 \le D/J <6$, to identify possible deviations
from that kind of transition. Indeed, significant
deviations from Ising universality have been observed for
$D/J \ge 5.9$, while for smaller values
of $D/J$ the simulational data are consistent with
an Ising--like transition. In addition, we identified and
located a line of compensation
points in the range $5.5 < D/J <6$. The main features of the
phase diagram are summarized in Fig. 7. The phase transition
line is based on analyzing various
quantities and taking into account finite--size
effects, as for the square lattice. Details of our
Monte Carlo findings will be discussed in the following.

The specific heat $C(T)$ shows for small and negative values of $D/J$ a
single maximum, giving rise to critical behaviour in the
thermodynamic limit. In case of an Ising--like
transition, its height is expected \cite{fis} to grow 
like $C_{max} \propto L^{\alpha/\nu}$ with the critical exponents
of the Ising universality class, $\alpha \approx 0.11$ and 
$\nu \approx 0.63$ \cite{pel}. Our simulational findings confirm
this scenario. As in the case of the
square lattice, upon increasing $D/J$, one encounters, eventually,
three maxima in $C(T)$, see Fig. 8. In complete analogy to
the two--dimensional case, the peak at the lower temperature, $T_l$,
is rather sharp and depends only very weakly on lattice size. It
signals the partial disordering of the B sublattice, with B spins being
flipped thermally from the ferromagnetic ('+' or '$-$') state
to 0. The maximum occurs at $k_BT_l/J \approx 0.6 (6-D/J)$. The
upper, rather broad maximum, at $T_u$, is non--critical
as well, stemming from dissolving the, at criticality still quite
large spin clusters on the A sublattice. $T_u$ is only very weakly 
affected by the strength of $D$, being
determined by the ferromagnetic coupling $J$. In between the two
non--critical maxima in $C(T)$, a critical peak shows up. It signals
the transition, at which both sublattice magnetizations vanish, with
quite pronounced local spin order on the A sublattice.

The type of the transition may be inferred from the size dependence 
of the critical peak, $C_{max}(L)$. Indeed, for single--ion
terms up to $D/J=5.8$, we find agreement with an Ising--type
transition, $\alpha/\nu \approx 0.17$. On further approach to the
degeneracy point, accurate Monte Carlo data with a fine
temperature resolution are required, due to the rather large
nonanalytic background term in $C$ and the sharpness of the
peak. In fact, other quantities may provide more easily and clearly
reliable clues on the type of transition for that part of the
transition line of the ferromagnetic phase.

\begin{figure} [h]
\vspace{1cm}
\begin{center}
\includegraphics[width=0.55\linewidth] {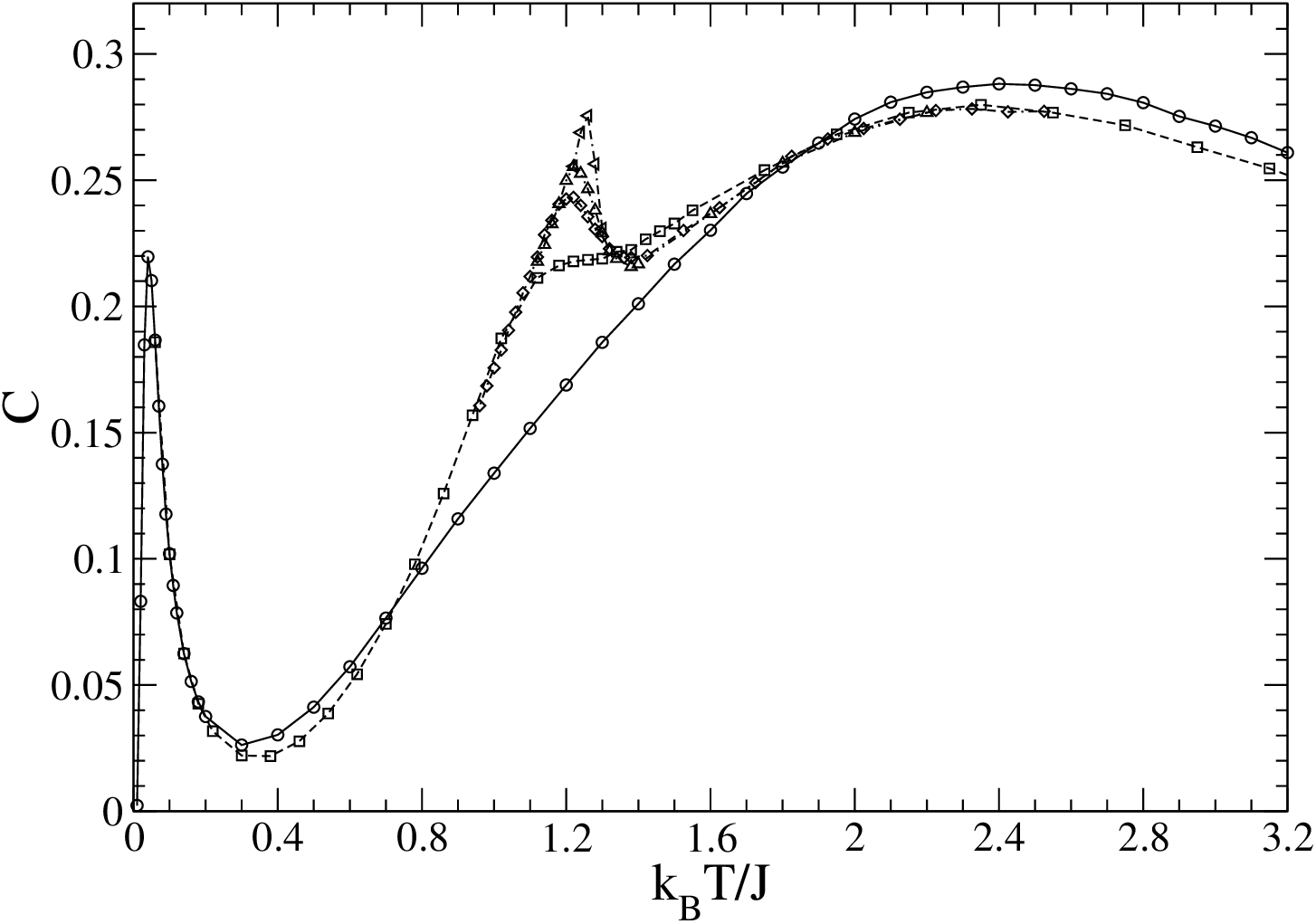}
\end{center}
\label{fig8}
\caption{Specific heat versus temperature for the model
    on the simple--cubic
    lattice at $D/J= 5.9$ for
    systems with $L$= 4(circles), 10 (squares), 16 (diamonds),
    20 (triangles up) and 32 (triangles left).} 
\end{figure}

Before discussing further the type of the phase transition
close to the degeneracy point, we shall deal with
the compensation points. Indeed, we identified such points 
in the range $5.5 < D/J < 6$. The resulting line is depicted
in Fig. 7. Two concrete examples are shown in Fig. 9, for
$D/J= 5.7$ and 5.9. As may be inferred from
that figure, the sublattice magnetization at the
compensation point decreases 
monotonically with decreasing single--ion term. Therefore, when
the compensation occurs at low magnetizations, the
accurate location of the compensation point is difficult, because
of strong finite--size effects in the critical region.  On the other
hand, with increasing $D/J$, the compensation point moves towards
lower temperatures, and finite size effects play usually no
significant role. In any event, in contrast to the two--dimensional
case, we find a line of compensation points for the simple--cubic
lattice. Obviously, the decrease in the magnetization of the
B sublattice, $|m_B|$, occurs in three dimensions
more drastically than for the square lattice, while $|m_A|$
changes there rather mildly in both cases.

\begin{figure} [h]
\vspace{1cm}
\begin{center}
\includegraphics[width=0.55\linewidth] {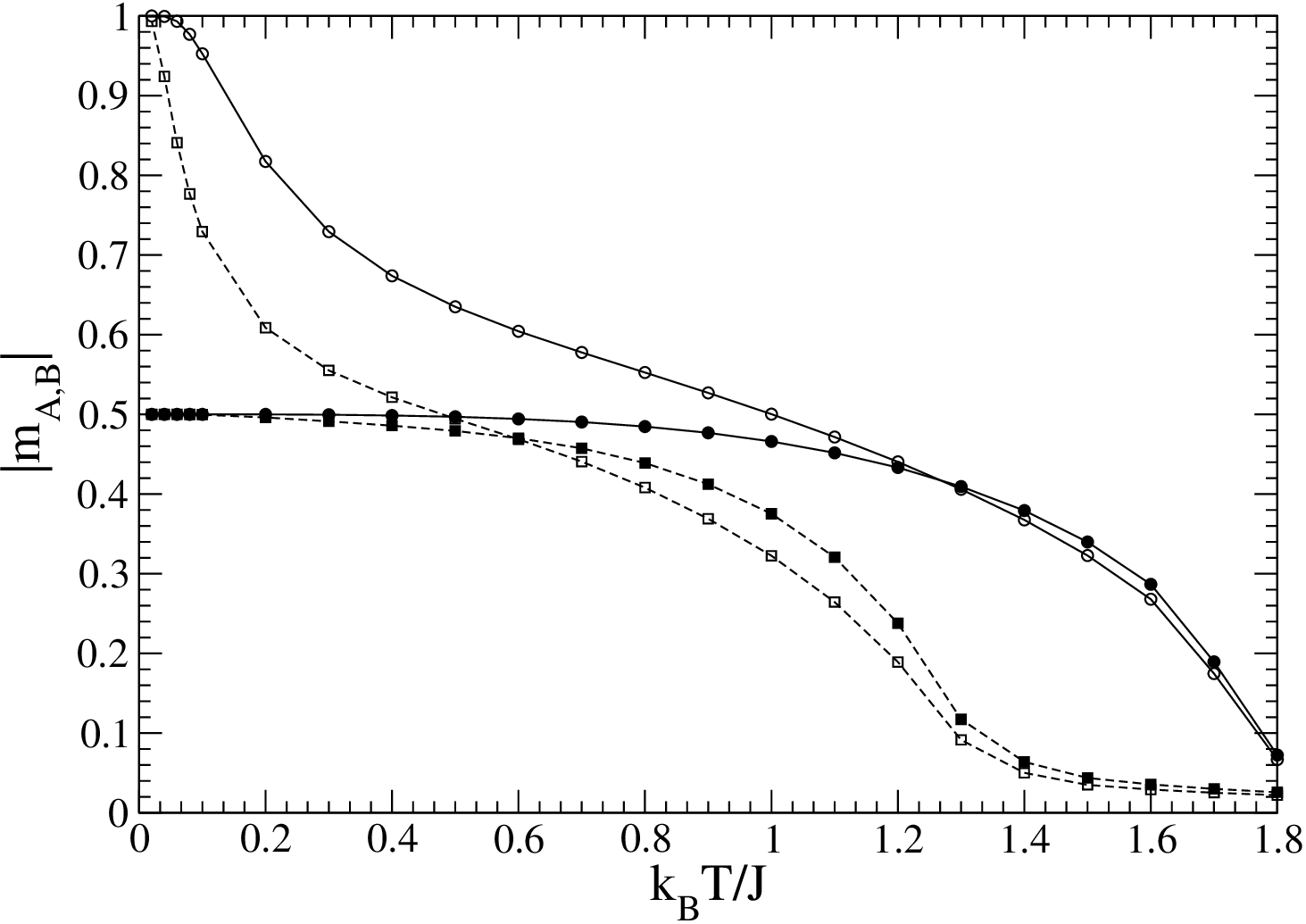}
\end{center}
\label{fig9}
\caption{Sublattice magnetizations $|m_A|$ and $|m_B|$ for the simple--cubic
    lattice with $L=20$ at $D/J= 5.7$ (circles) and 5.9 (squares).} 
\end{figure}

Let us now turn back to the discussion on the type of
phase transition. For $D/J \le 5.8$, the data on the
susceptibilty $\chi$ confirm the Ising--like character of
the transition. In particular, the size dependence
of the height of the maximum in $\chi$, $\chi_{max}(L)$, is found to be
consistent with Ising
criticality, $\chi_{max} \propto L^{\gamma/\nu}$, where
$\gamma \approx 1.24$ and $\nu \approx 0.63$, thus 
$\gamma/\nu \approx 1.97$. Indeed, from our
simulational data we obtain characteristic exponents
close to 2. However, at $D/J$= 5.9, we observe, for systems
sizes ranging from $L= 8$ to $L= 32$, a substantially lower (effective)
exponent, of about 1.7. Because the peak in $\chi$ gets extremely
sharp, very accurate simulational data with a very fine temperature mesh are
needed to arrive at safe conclusions. A more convenient way to 
monitor the possible change in the type of the transition will
be discussed below. 

\begin{figure}[h]
\vspace{1cm}
\hspace{-1.2cm}
\begin{center}
\includegraphics[width=0.44\linewidth]{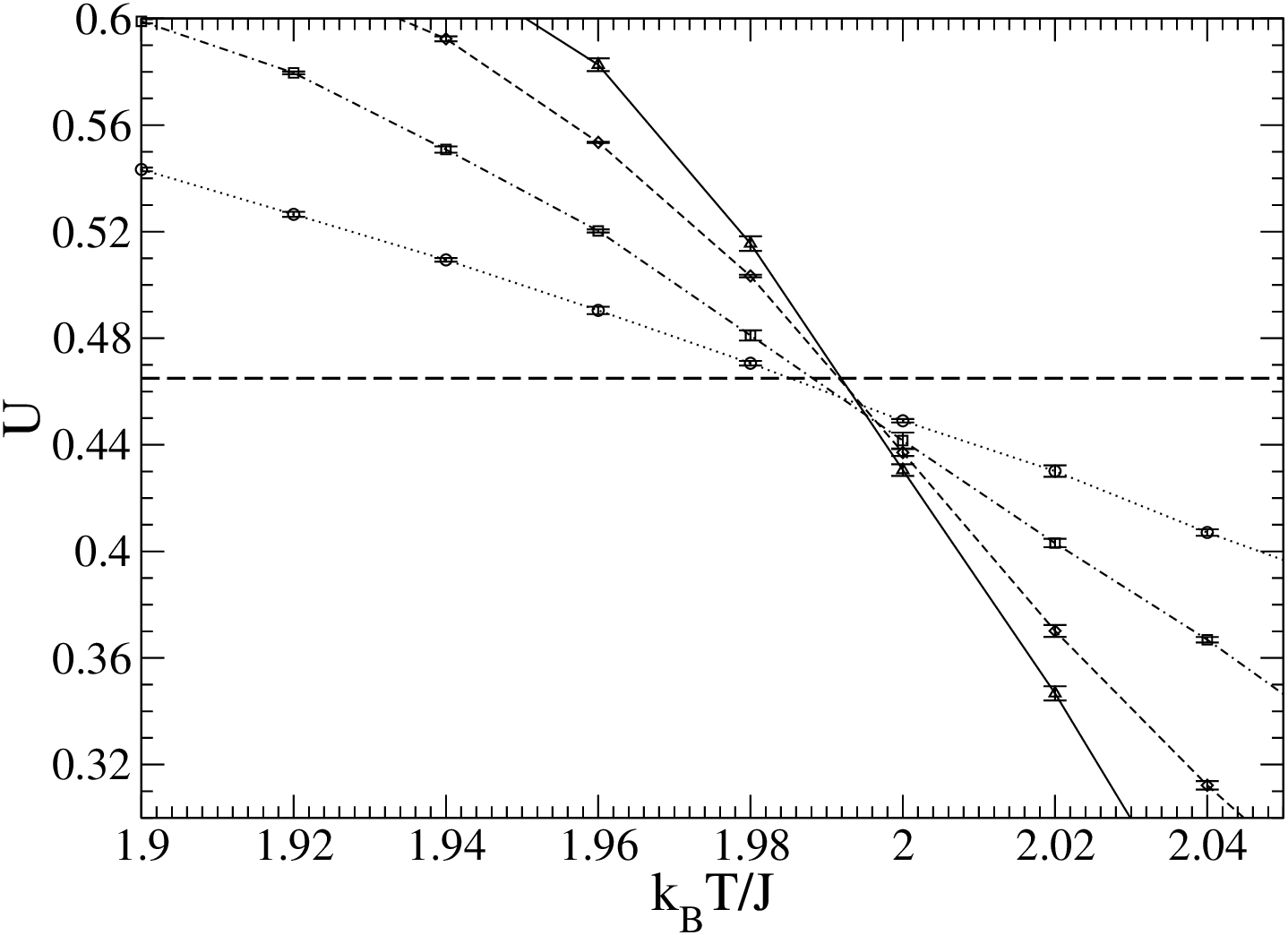}
\hspace{0.7cm}
\includegraphics[width=0.44\linewidth]{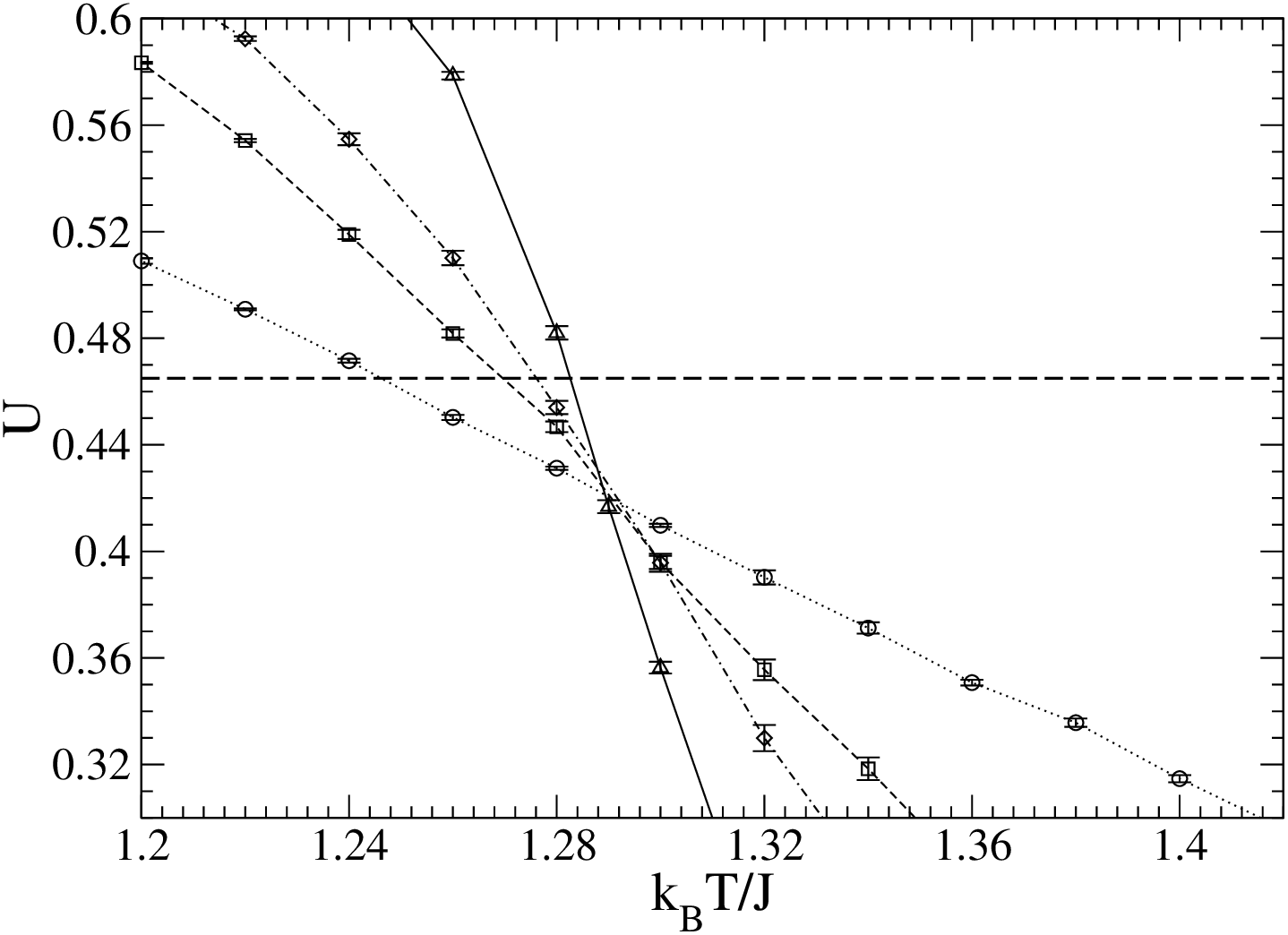}
\end{center}
\label{fig10}
\caption{(a) Left: Binder cumulant $U$ versus temperature 
  at $D/J= 5.5$ for lattices with $L$= 8 (circles), 12 (squares),
  16 (diamonds) and 20 (triangles). (b) Right: $U$ 
  versus temperature at $D/J= 5.9$
  for lattices with $L$= 10 (circles), 16 (squares), 20 (diamonds),
  and 32 (triangles). The horizontal lines indicate the critical
  Binder cumulant of an isotropic three--dimensional Ising model
  in the thermodynamic limit \cite{has}.}
\end{figure}

Interestingly, our analysis of the Binder cumulant $U$ seems to indicate
substantial deviations from an Ising--like transition at about $D/J \approx
5.9$ as well. For smaller values of $D/J$ the intersection values of
the cumulant curves for different system sizes, already for
fairly small systems, seem to agree
with the expected asymptotic value of the critical Binder
cumulant for isotropic Ising systems \cite{has}, $U^* \approx 0.465$. An
example is depicted in Fig. 10a, for $D/J =5.5$, with the intersection
points, for the simulated finite lattices, approaching the
asymptotic value from below, when increasing
the system size. At larger single--ion anisotropy, $D/J \ge 5.9$, the
intersection points of the curves are appreciably lower than $U^*$,
as shown in Fig. 10b for $D/J= 5.9$. However, it is not completely
clear, whether the tendency reflects stronger finite--size effects or 
a change in the type of the phase transition.

To get more evidence for a possible change of the nature of
the transition, the histograms
for the magnetization, $p(m)$, turned out to be most
instructive. Already for small lattices, $L= 4$, one sees, close
to the transition, a qualitative change
of the histograms. We did simulations close to the
transitions in the
range $5.85 \ge D/J \ge 5.98$, using an increment of 0.01. We
observe a dramatic change in the form of the histograms
around $D/J \approx 5.91$. Below that value, there
is no central peak and thus no indication
of phase coexistence when crossing the transition, in
contrast to the situation closer to the degeneracy point, where
a central peak, in addition to the symmetric peaks
at $\pm m_0$, indicates coexistence of the ordered and
disordered phases and, accordingly, a transition
of first order. That distinction
persists for larger system sizes. Examples are
displayed in Figs. 11 a, for $D/J= 5.85$, and 11 b, for
$D/J= 5.975$. Based on these observations, we may
tentatively locate the tricritical point at $D/J= 5.91 \pm 0.03$. Note
that such a change in the form of the histograms does not occur in two
dimensions, as has been discussed above, see also Fig. 6.

\begin{figure}[h]
\vspace{1cm}
\hspace{-1.2cm}
\begin{center}
\includegraphics[width=0.44\linewidth]{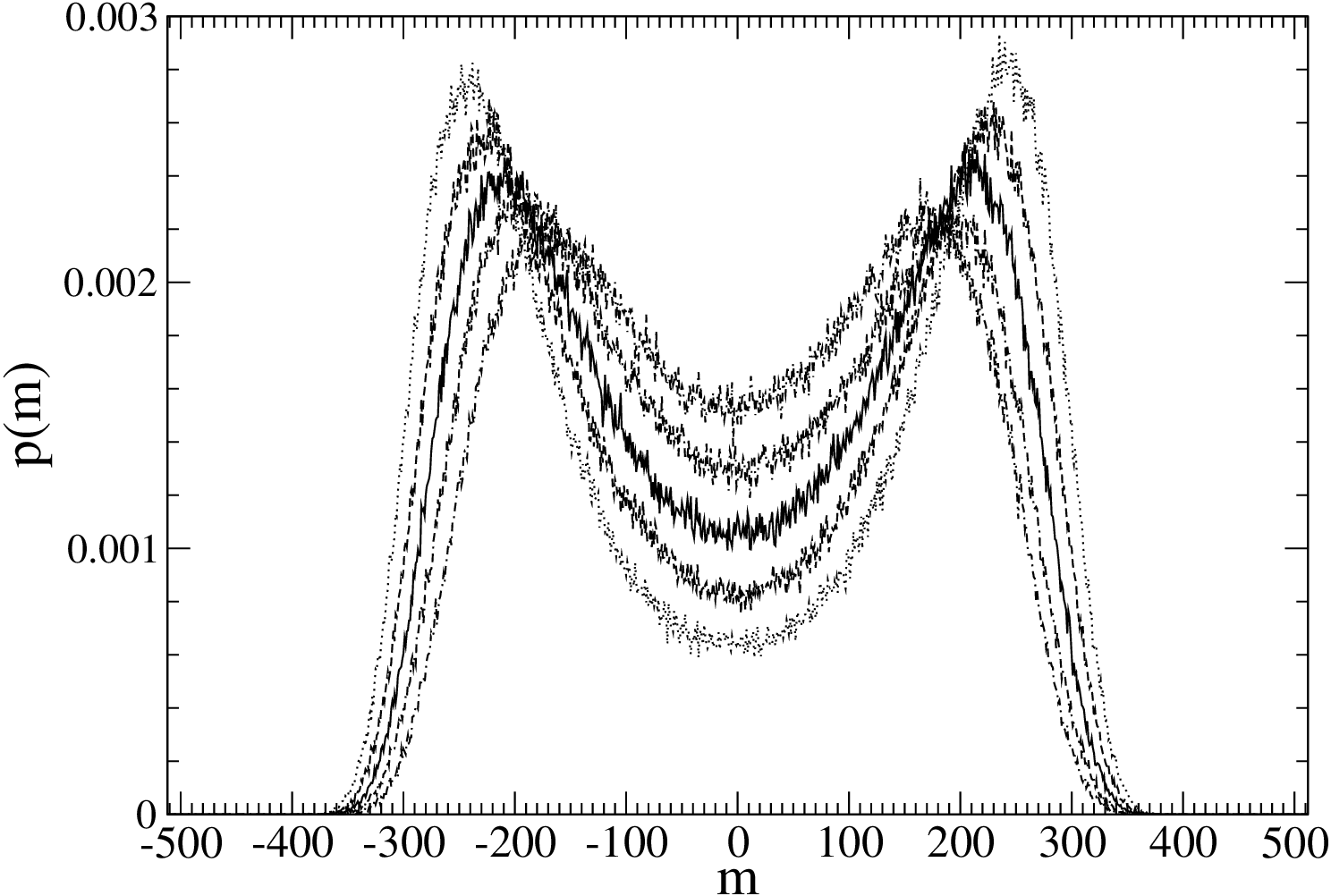}
\hspace{0.7cm}
\includegraphics[width=0.44\linewidth]{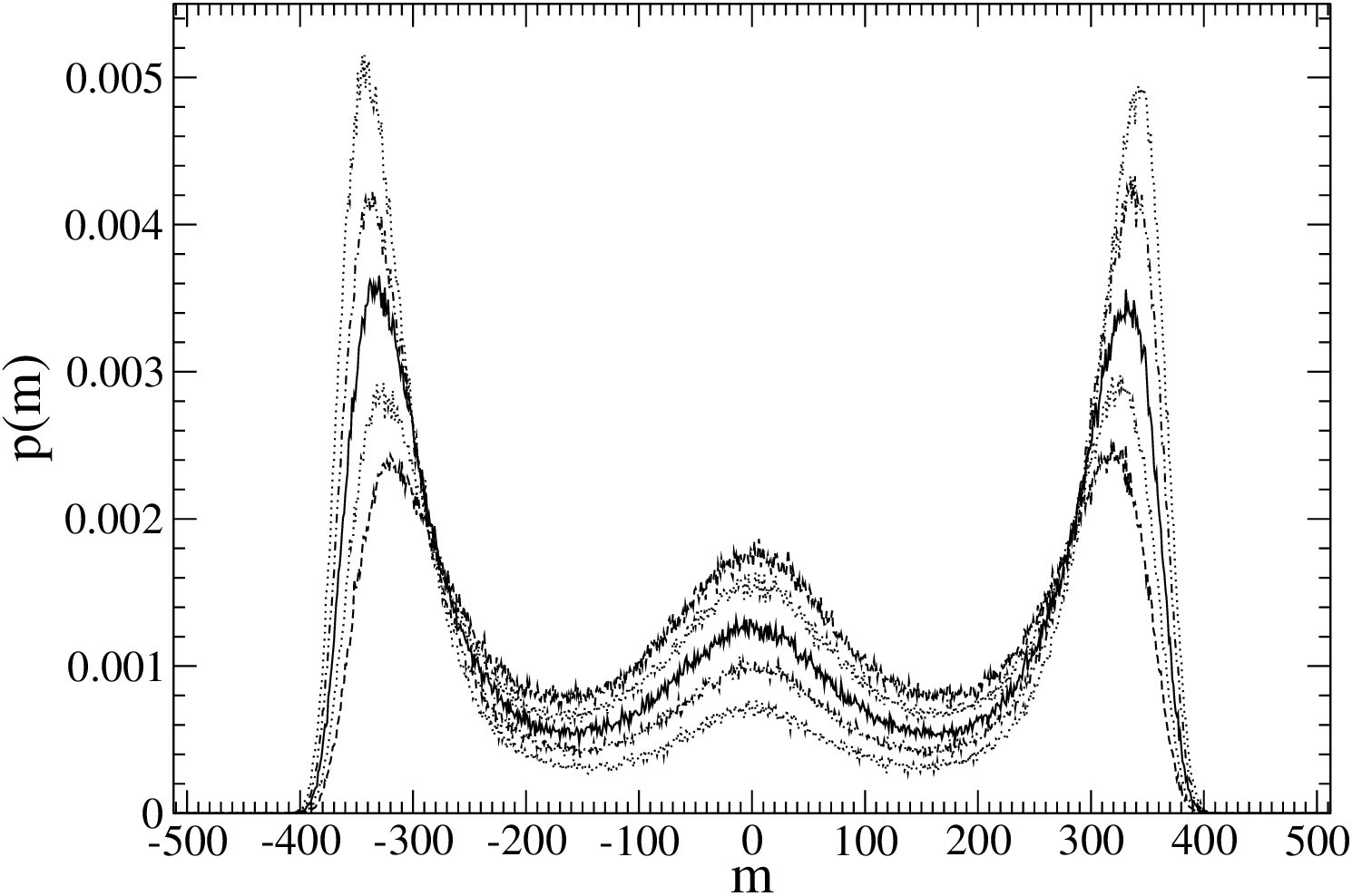}
\end{center}
\label{fig11}
\caption{(a) Left: Histogram of the total magnetization, $p(m)$,
  for $L$=8 and $D/J= 5.85$, at temperatures crossing the
  transition, $k_BT/J$= 1.32, 1.36, 1.40, 1.44, and 1.48, where the
  maxima move towards the center with increasing
  temperature. (b) Right: $p(m)$
  for $L$= 8 and $D/J= 5.975$, at temperatures crossing the
  transition, $k_BT/J$= 0.47, 0.51, 0.55, 0.59, and 0.63, where the
  central peak grows in height with increasing temperature.}
\end{figure}

In summary, the present analysis on the mixed--spin model
on a simple--cubic lattice shows clearly a line
of compensation points, and allows to locate approximately the
tricritical point.

\section{Summary}
\label{sec4}

We have studied a mixed--spin Ising model with
ferromagnetic couplings, $J$, between spins 1/2 and 1
on neighbouring sites of square and
simple--cubic lattices, the two types of spins forming
a bipartite lattice. An additional quadratic single--ion
term, $D$, acts upon the S=1 spins. We mainly used
standard Monte Carlo simulations to compute various
thermodynamic properties as well as the Binder cumulants and
histograms of the total magnetization.

The model on the square lattice has been shown
to display a continuous phase transition
of Ising--type, presumably up to the degeneracy point at
$D/J$ =4. No
compensation point has been found. Close to the
degeneracy point, the model displays an intriguing three--peak
structure in the specific heat as a function of temperature. The 
sharp, but non--critical anomaly at low temperatures arises
from flipping S=1 spins into the state 0,
while the broad non--critical maximum at high
temperatures stems from thermal
activation of spins in fairly large clusters of S=1/2 spins
persisting above the phase transition. At temperatures in between, the
critical peak shows up. Both anomalies  
may cause difficulties in low-- and high--temperature expansions,
which have predicted, incorrectly, the existence of a tricritical
point. The suggestion on the absence of a compensation point has been
confirmed, albeit the magnetization on the S=1 sublattice decreases
rapidly near the anomaly of the specific heat at low temperatures.

In the case of the simple--cubic lattice, the specific heat
displays a similar three--peak structure, with two
non--critical maxima and the critical peak in between. Sufficiently
far away from the degeneracy point, the ferromagnetic phase
disorders via a continuous, Ising--like transition. In the
vicinity of the degeneracy point, $D/J= 6$, this transition seems
to be of first
order. The evidence for that kind of transition 
is mainly based on the type of the
histograms of the magnetization, showing
phase coexistence. We tentatively
locate the tricritical point at $D/J= 5.91 \pm 0.03$. In
addition, we determined a line of compensation
points, arising from the degeneracy point. Thus, in three
dimensions, the mean--field theory appears to give at
least qualitatively correct predictions. However, in two
dimensions the mean--field theory is 
found to be incorrect, even qualitatively.

\vspace{1cm}

\section {Acknowledgement}

W.S. thanks the School of Physics at the University of New 
South Wales for the very kind hospitality during his stay there, which
was supported by the Gordon Godfrey Fund.

\end{document}